\documentclass[prl,twocolumn,showpacs,floatfix,amsmath,amssymb,superscriptaddress]{revtex4-1}
\usepackage{amsfonts}
\usepackage{stmaryrd}
\usepackage{bbm}
\usepackage{mathrsfs}
\usepackage{tipa}
\usepackage{amssymb}
\usepackage{txfonts}
\usepackage{graphicx}
\usepackage{dcolumn}
\usepackage{epstopdf}
\usepackage[colorlinks,linkcolor=blue,urlcolor=blue,citecolor=blue]{hyperref}
\usepackage{multirow}
\usepackage{subfigure}
\usepackage{url}

\begin{document}
\newcommand*{\cm}{cm$^{-1}$\,}
\newcommand*{\TC}{T$_c$\,}
\newcommand*{\TCDW}{T$_{CDW}$\,}
\newcommand*{\A}{2$\Delta$/k$_{B}$T$_{CDW}$\,}
\newcommand*{\Sr}{Sr$_{3}$Rh$_{4}$Sn$_{13}$\,}
\newcommand*{\Ca}{(Sr$_{0.5}$Ca$_{0.5}$)$_{3}$Rh$_{4}$Sn$_{13}$\,}
\newcommand*{\SrCa}{(Sr$_{x}$Ca$_{1-x}$)$_{3}$Rh$_{4}$Sn$_{13}$\,}

\title{Revealing correlation effect of Co 3d electrons in La$_{3}$Co$_{4}$Sn$_{13}$ and Ce$_{3}$Co$_{4}$Sn$_{13}$ by infrared spectroscopy study}
\author{W. J. Ban}
\affiliation{Institute of Physics, Chinese Academy of Sciences, Beijing, China}

\author{J. L. Luo}
\affiliation{Institute of Physics, Chinese Academy of Sciences, Beijing, China}

\author{N. L. Wang}
\email{nlwang@pku.edu.cn}
\affiliation{International Center for Quantum Materials, School of Physics, Peking University, Beijing 100871, China}
\affiliation{Collaborative Innovation Center of Quantum Matter, Beijing, China}

\begin{abstract}
We report resistivity, specific heat and optical spectroscopy measurements on single crystal samples of La$_{3}$Co$_{4}$Sn$_{13}$ and Ce$_{3}$Co$_{4}$Sn$_{13}$. We observed clear temperature-induced spectral weight suppression below 4000 \cm for both compounds in conductivity spectra $\sigma_1(\omega)$, indicating the progressive formation of gap-like features with decreasing temperature. The suppressed spectral weight transfers mostly to the higher energy region. The observation reflects the presence of correlation effect in the compounds. We attribute the correlation effect to the Co 3d electrons.

\end{abstract}

\pacs{}

\maketitle\emph{}
\section{introduction}

Intermetallic compounds  with composition A$_{3}$T$_{4}$X$_{13}$ (where A stands for Sr, Ca or rare earth element, T for a transition metal and X for Ge or Sn) crystallizing in Yb$_{3}$Rh$_{4}$Sn$_{13}$ type structure have attracted high attention due to the existence of rich electronic and magnetic phenomena. The structure contains two inequivalent X sites (X1 and X2). The X1, R, T, and X2 atoms occupy respectively the 2a, 6d, 8e, and 24k sites of the cubic structure with Pm\={3}n space group (No. 223). The X1 atoms locate the origin of the unit cell and $(X1)(X2)_{12}$ form an edge-sharing icosahedra, while the T atom and six X2 atoms form the $T(X2)_{6}$ trigonal prisms. Superconductivity, charge density wave, complex magnetism, heavy fermions as well as mixed-valence behaviors have been found in this class of materials\cite{doi:10.1080/02603590600666215,C7CE00419B,REMEIKA1980923,Israel2005251,Sato1993630,PhysRevB.56.8346,LyleThomas20061642,0953-8984-19-38-386207,0953-8984-20-39-395208,doi:10.1143/JPSJ.79.113705,PhysRevB.83.184509,PhysRevB.85.205120,PhysRevB.86.024522,PhysRevB.86.064504,PhysRevLett.109.237008,Rai2015Intermediate,PhysRevB.93.035101,doi:10.1021/cm504658h}. Many of the Sn-based compounds, for example A$_{3}$T$_{4}$Sn$_{13}$ (A=La, Sr, Ca and T=Rh and Ir), show interesting coexistence of structural phase transition and superconductivity. The structural phase transition with formation of superlattice modulation was suggested to arise from charge density wave (CDW) instability. Our recent optical spectroscopy study on Sr$_{3}$Ir$_{4}$Sn$_{13}$ and (Sr$_{1-x}$Ca$_{x}$)$_{3}$Rh$_{4}$Sn$_{13}$ (x=0 and 0.5) revealed formation of density wave-like energy gaps in optical conductivity below the structural phase transition temperatures \cite{PhysRevB.90.035115,arXiv:1609.04206}. Notably, although the structural phase transition leads prominent peak in the specific heat and sudden change in resistivity and magnetic susceptibility, the energy gap structure in optical measurement is rather weak if compared with the quasi one- or two-dimensional CDW compounds.

On the other hand, the rare-earth element containing compounds of A$_{3}$T$_{4}$Sn$_{13}$ (A=rare earth element) often exhibit complex magnetic and heavy fermion properties. For example, Ce$_{3}$Co$_{4}$Sn$_{13}$ is known as a non-superconducting heavy fermion dense Kondo system with a Kondo temperature of about 1.5 K. The compound exhibits a broad peak in the electronic specific heat coefficient \emph{C}(T)/T at about 0.7 K with the maximum value of about 4 J/(K$^2$mole) which is due to the short-range magnetic correlation of Ce 4f electrons \cite{ISI:000340983800029}. Besides Ce 4f electrons, the Co 3d electrons in this compound also play roles in its physical properties. Unlike the 5d or 4d electrons from Ir or Rh atoms, the Co 3d orbitals are more localized in real space, leading to sizeable correlation effect. A density functional calculations of R$_{3}$Co$_{4}$Sn$_{13}$ indicates a strong hybridization between Co 3d and Sn2 5p states near the $E_{F}$. Under the Co(Sn2)$_6$ trigonal prisms in the crystal structure, the 3d orbitals of Co are split into $d_{z^{2}}$, degenerate $d_{xy}(d_{x^{2}+y^{2}})$, and $d_{xz}(d_{yz})$ orbitals from low to high energy. The two antibonding bands mixed by Co $d_{xz}(d_{yz})$ and Sn2 5p cross the $E_{F}$ to form a relatively broad conduction band and the complicated Fermi surface (FS). The $d_{z^{2}}$ and degenerate $d_{xy}(d_{x^{2}+y^{2}})$ bands are localized \cite{PhysRevB.79.094424}. An x-ray photoemission spectroscopy study in combination with band structure calculations yield evidence for the d-electron correlation in La$_{3}$Co$_{4}$Sn$_{13}$ and Ce$_{3}$Co$_{4}$Sn$_{13}$ compounds \cite{PhysRevB.91.035101}.

In order to reveal the possible correlation effect of 3d transition metal electrons in such intermetallic system, it is of high interest to study the properties of Co-based compounds and compare them with Rh and Ir based compounds in such structure family. In this work, we present resistivity, specific heat and optical spectroscopy measurements on single crystal samples of La$_{3}$Co$_{4}$Sn$_{13}$ and Ce$_{3}$Co$_{4}$Sn$_{13}$. The resistivity and specific heat measurements revealed weak anomalies near 150 K, which should be driven by CDW instabilities. However, no visible suppression associated with formation of CDW energy gap could be identified. Instead, we found a clear temperature-dependent spectral weight suppression below about 4000 \cm in conductivity spectra $\sigma_1(\omega)$ for both compounds. The suppressed spectral weight is transferred to the higher energy region. We elaborate that this suppression is caused by the correlation effect from Co 3d electrons.

\section{EXPERIMENTAL}

The La$_{3}$Co$_{4}$Sn$_{13}$ and Ce$_{3}$Co$_{4}$Sn$_{13}$ single crystals were grown by Sn self-flux method. High-purity elements were mixed in the molar ratio of La:Co:Sn=1:1:20 and sealed in an evacuated quartz tube. The tube was heated up to 1273 K, dwelled for 5 h, and then cooled to 773 K over 250 h. Large pieces of single crystals with shiny surfaces were yielded by eliminating excess Sn flux through a centrifugal machine. The resulting crystals have dimensions of several millimeters. Phase examination of the crystals were studied by X-ray diffraction (XRD) on a Rigaku Smartlab High Resolution Diffractometer at room temperature using Cu K $\alpha_{1}$ radiation ($\lambda=1.5406 {\AA}$). The dc resistivity were measured by a four-probe method, and the electrical current flows along the direction parallel to the (110) plane of the crystal. The specific heat was measured by a relaxation-time method. Both measurements were conducted on a commercial Quantum Design Physical Properties Measurement System (PPMS).

The optical reflectance measurements were performed on as-grown shinny surfaces of the single crystals which were identified to be the (110) plane by XRD measurements with a combination of Bruker 113v, Vertex 80v and and grating-type spectrometers in the frequency range from 40 to 44000 \cm. An in-situ gold and aluminum over-coating technique was used to get the reflectance \cite{Homes:93}. The measured reflectance was then corrected by multiplying the available curves of gold and aluminum reflectivity at different temperatures. The real part of conductivity $\sigma_1(\omega)$ was obtained by the Kramers-Kronig transformation of R($\omega$) employing an extrapolation method with X-ray atomic scattering functions \cite{PhysRevB.91.035123}.

\section{RESULTS AND DISCUSSIONS}

\begin{figure}[htbp]
  \centering
  \includegraphics[width=7.5cm]{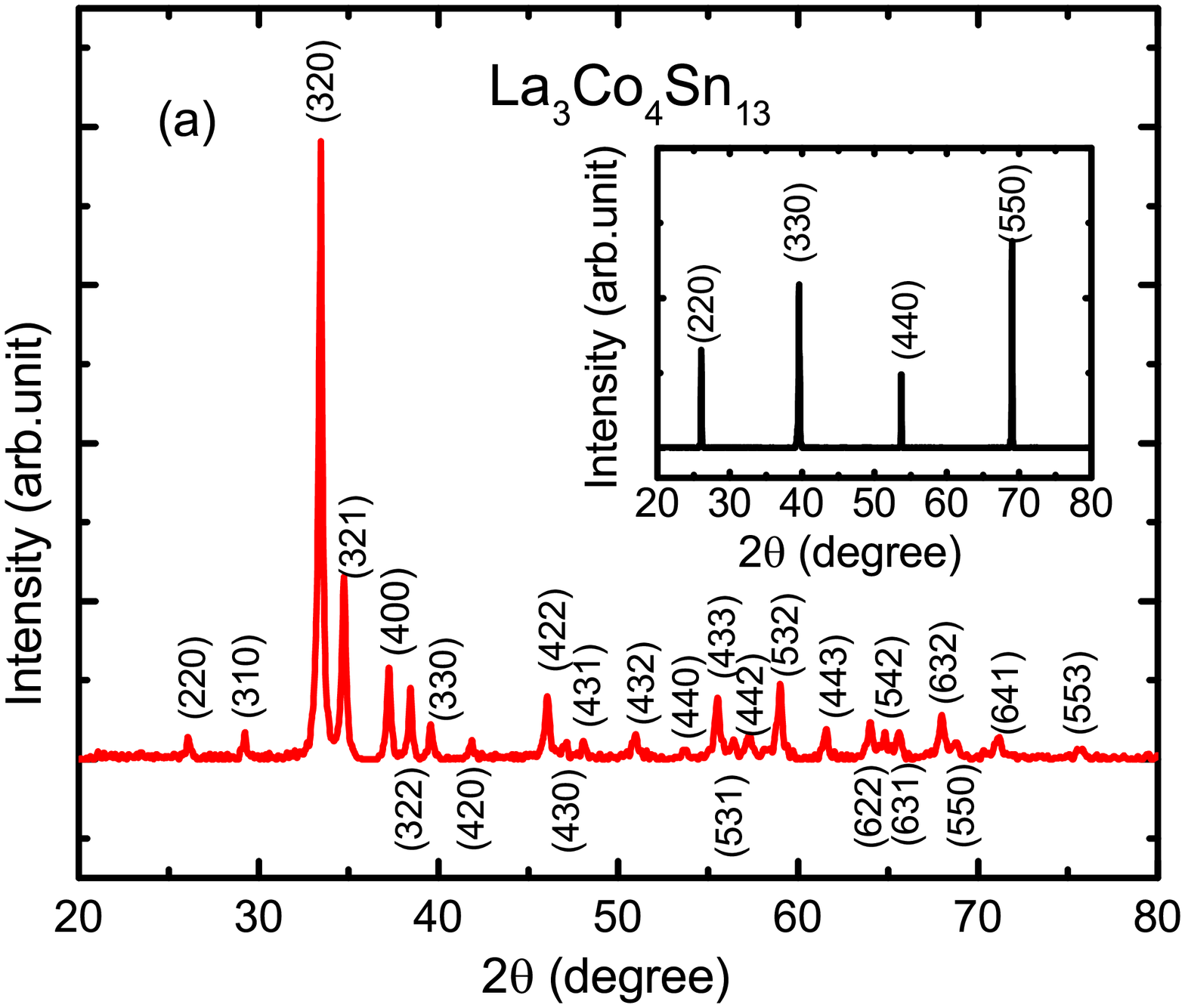}\\
  \includegraphics[width=7.5cm]{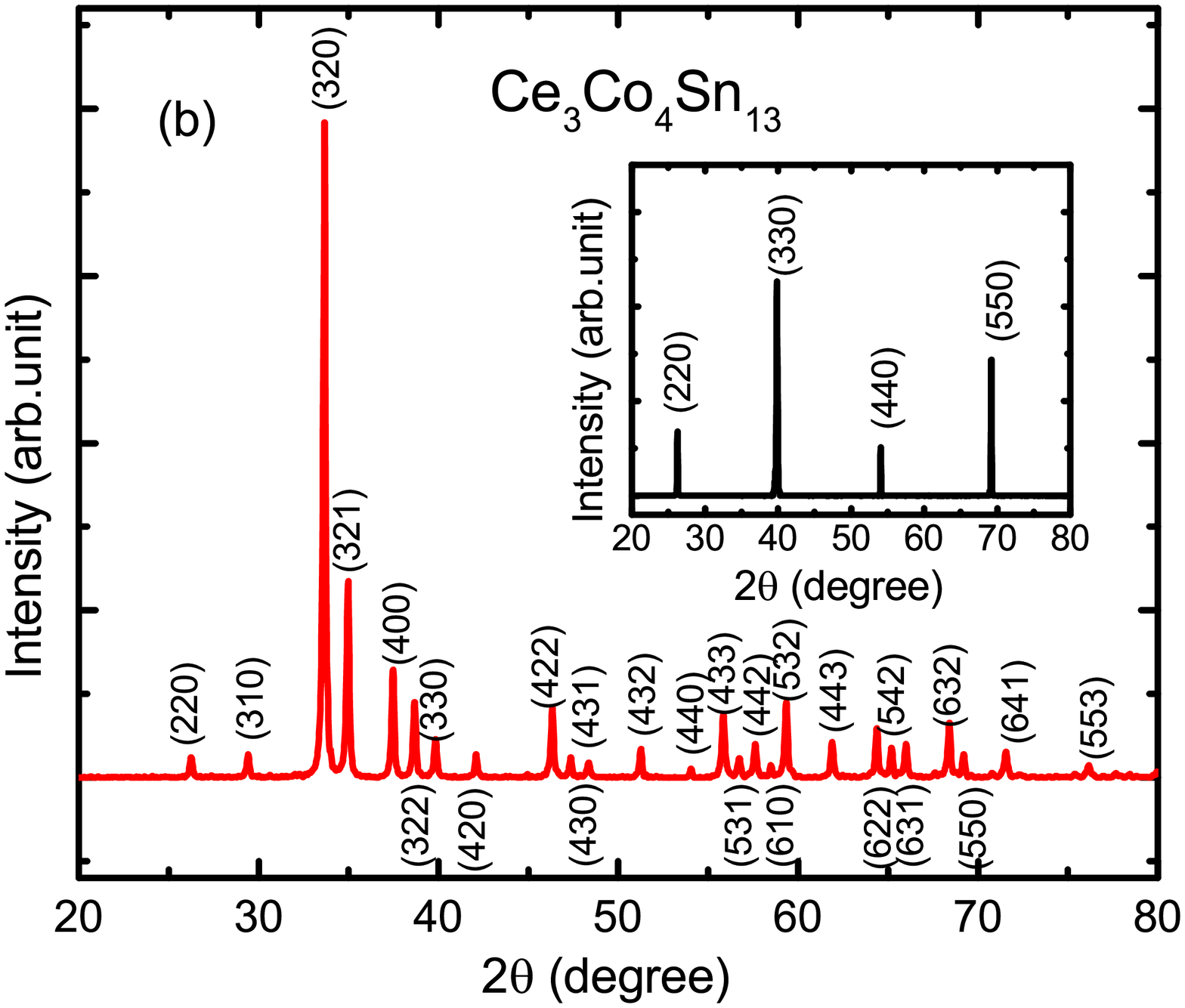}\\
    \caption{Powder XRD patterns at room temperature for the sample La$_{3}$Co$_{4}$Sn$_{13}$  (a) and Ce$_{3}$Co$_{4}$Sn$_{13}$ (b). Inset: the XRD patterns measured on the (110) surfaces for the two samples. Reflections are indexed with respect to the Yb3Rh4Sn13-type structure (space group Pm3n).}\label{Fig:xrd}
\end{figure}

Figure \ref{Fig:xrd} show the powder XRD patterns and the insets show the XRD patterns measured on the (110) surfaces at room temperature for the samples of La$_{3}$Co$_{4}$Sn$_{13}$ and Ce$_{3}$Co$_{4}$Sn$_{13}$,  which are consistent with the $Yb_{3}Rh_{4}Sn_{13}$-
type $Pm\bar{3}n$ space group as in previous studies\cite{PhysRevB.88.115113,PhysRevB.86.205113,PhysRevB.88.155122,ISI:000369375800046,lebarski2014199,PhysRevB.94.075109}. For La$_{3}$Co$_{4}$Sn$_{13}$ and Ce$_{3}$Co$_{4}$Sn$_{13}$ sample, the Lattice parameters obtained from the XRD datas are a=9.6369 and 9.5934 ${\AA}$, respectively, close to the previously reported value\cite{PhysRevB.88.115113,PhysRevB.86.205113,PhysRevB.88.155122,ISI:000369375800046,lebarski2014199,PhysRevB.94.075109,PhysRevB.85.205120,PhysRevB.79.094424,Israel2005251,ISI:000369375800046,PhysRevB.88.155122,LyleThomas20061642}, indicating the good quality of our samples.

\begin{figure}[htbp]
  \centering
  \includegraphics[width=7.5cm]{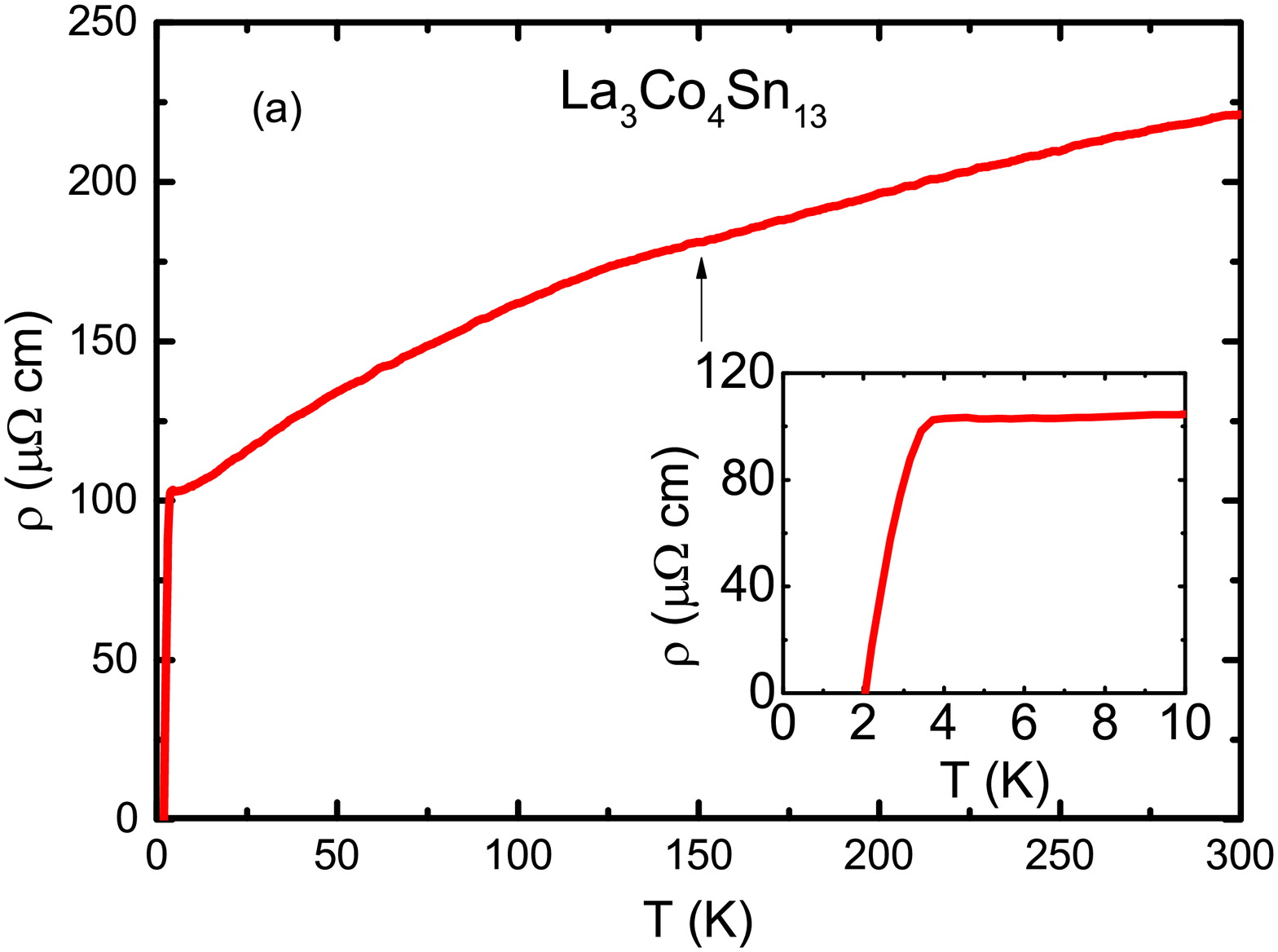}\\
  \includegraphics[width=7.5cm]{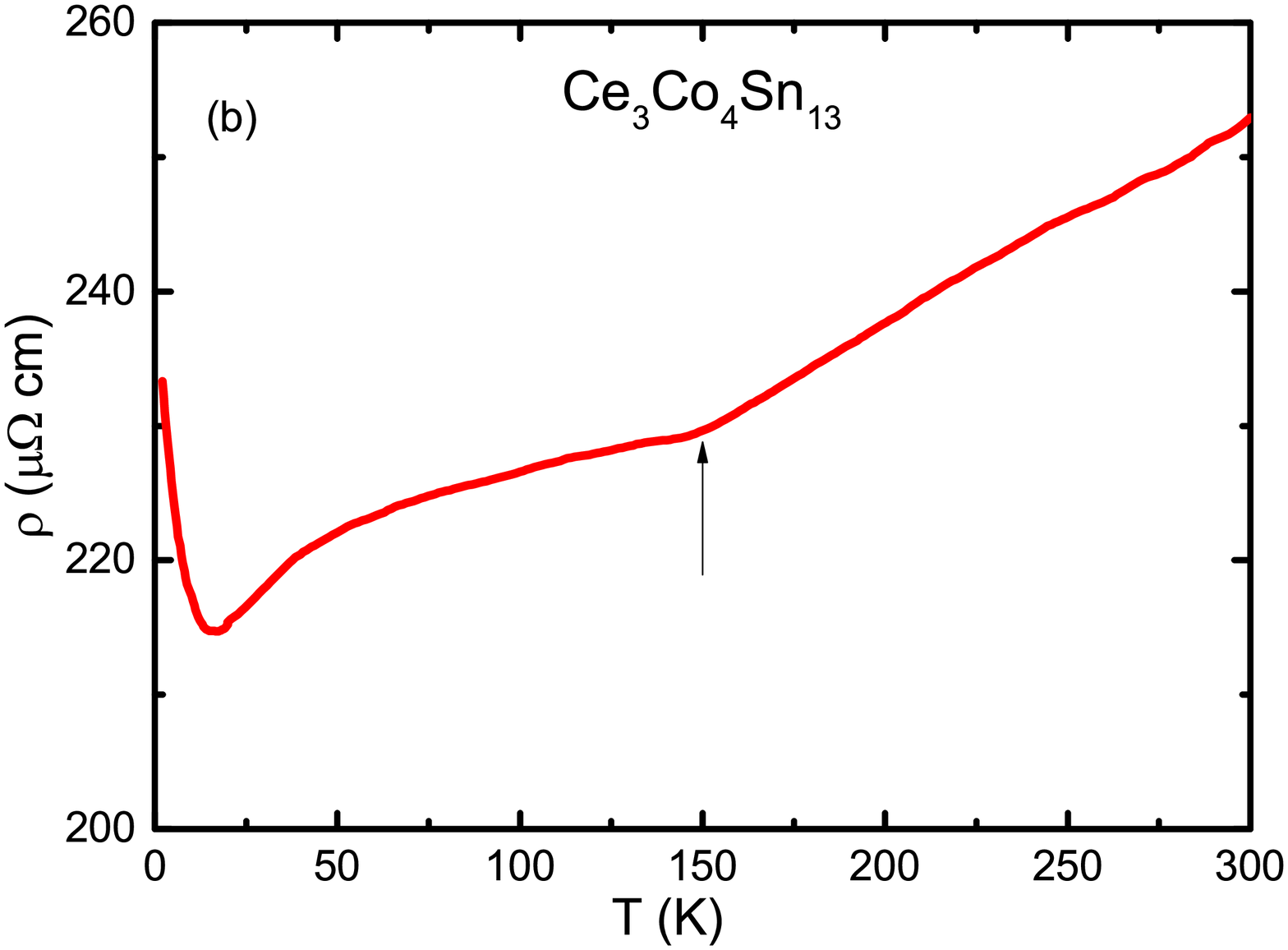}\\
    \caption{The temperature dependent resistivity $\rho(T)$ of La$_{3}$Co$_{4}$Sn$_{13}$  (a) and Ce$_{3}$Co$_{4}$Sn$_{13}$ (b).  The inset in (a) shows the low-temperature portion of the $\rho(T)$ to depict the sharp transition to the superconducting state.}\label{Fig:resistivity}
\end{figure}

Figure \ref{Fig:resistivity} presents the temperature dependence of resistivity for La$_{3}$Co$_{4}$Sn$_{13}$ and Ce$_{3}$Co$_{4}$Sn$_{13}$ single crystals. For La$_{3}$Co$_{4}$Sn$_{13}$, metallic temperature dependence is observed in the whole temperature range. A very weak kink appear near 150 K. The observations are similar to the published data in the literature\cite{LyleThomas20061642,PhysRevB.88.115113}. Cheung et al. suggested that $T^{*}$ is associated with a second-order structural transition with q=(0.5, 0.5, 0)\cite{PhysRevB.93.241112}. Lue et al. suggested that the observed phase transition behavior at $T^{*}$ can be associated with the partially gapped Fermi surfaces which would be appropriate for its isostructural analog of $Sr_{3}Ir_{4}Sn_{13}$, which has been claimed to possess charge-density-wave (CDW) behavior \cite{PhysRevB.88.115113}. At very low temperature, the sample shows a superconducting transition. The resistivity begin to drop at about 3.8 K and reaches zero at 2 K. For Ce$_{3}$Co$_{4}$Sn$_{13}$, upon cooling the resistivity decreases continuously until 15K. Below 15K, the resistivity exhibits $\rho\sim-lnT$ behavior. The behavior was explained as due to the Kondo impurity mechanism\cite{ISI:000340983800029}. Comparing with La$_{3}$Co$_{4}$Sn$_{13}$, a more distinct anomaly is observed at $T^{*}\approx150$ K. The observations are similar to the published data in the literature\cite{PhysRevB.94.075109,LyleThomas20061642,PhysRevB.88.115113,PhysRevB.85.205120,ISI:000340983800029,ISI:000247666000026,ISI:000238426600041,PhysRevB.86.205113}. A recent studies of Ce3Co4Sn13 explained this distinct anomaly behavior as a result of a possible local
distortion of the trigonal Sn2 prisms around Co, which could change
the electronic structure near the Fermi level\cite{PhysRevB.88.155122,LyleThomas20061642,PhysRevB.86.205113}. Lue et al. suggested that the observed phase transition behavior can be related to an electronic origin and/or electron-lattice coupling such as the formation of a CDW\cite{PhysRevB.85.205120}.

\begin{figure}[htbp]
  \centering
  \includegraphics[width=7.5cm]{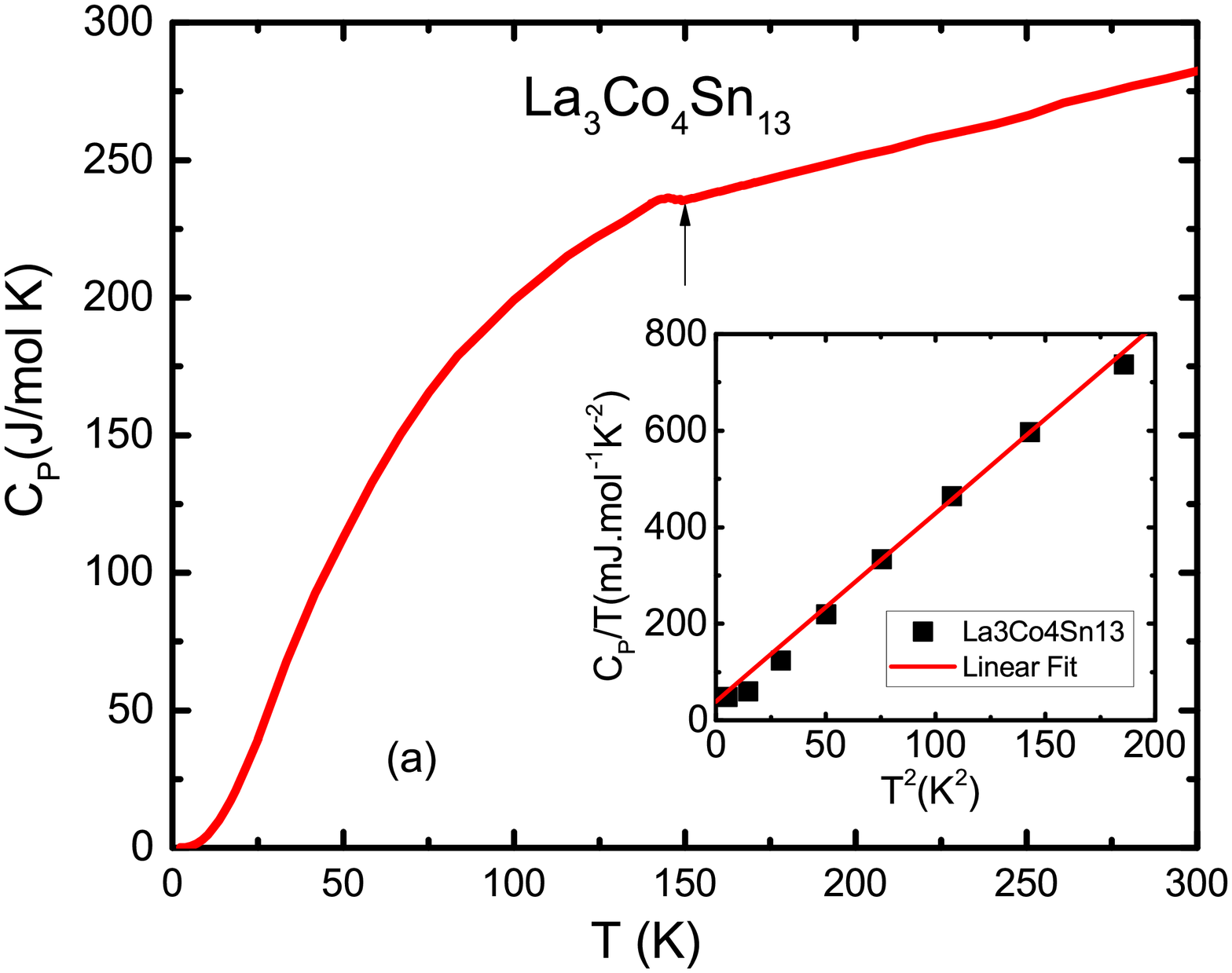}\\
  \includegraphics[width=7.5cm]{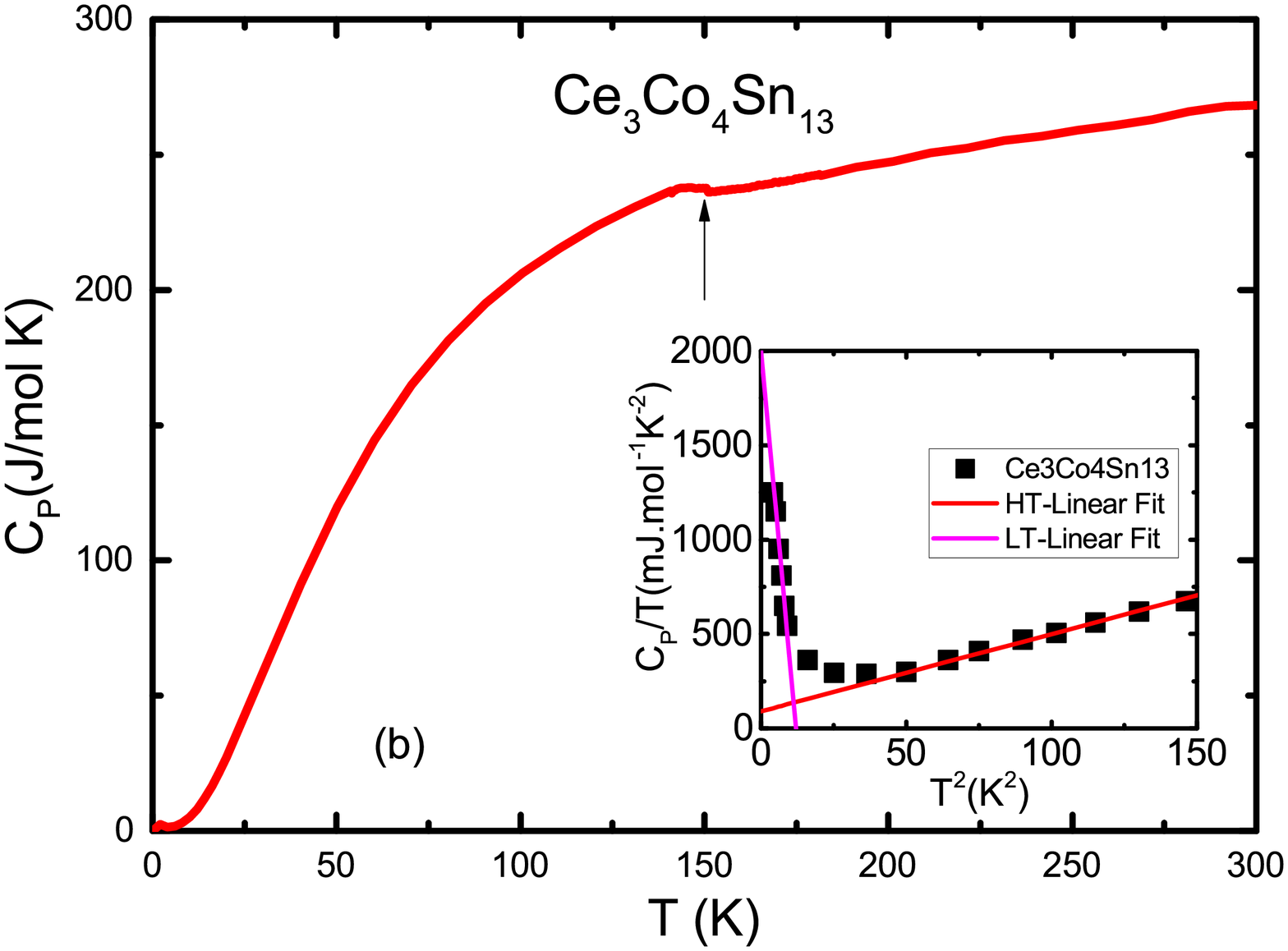}\\
    \caption{The temperature dependent specific heat $C_{P}(T)$ of La$_{3}$Co$_{4}$Sn$_{13}$  (a) and Ce$_{3}$Co$_{4}$Sn$_{13}$ (b).  The inset  is a plot of $C_{P}/T$ as a function of $T^{2}$ together with a fit to a straight line at the low temperature. And the additional pink line shows the fit which considering the effect of c-f hybridization at even low temperature}\label{Fig:HC}
\end{figure}

Figure \ref{Fig:HC} shows the temperature dependent specific heat for the two samples. The specific heat $C_{P}(T)$ also reveals an anomaly at corresponding $T^{*}$. However, the feature in La$_{3}$Co$_{4}$Sn$_{13}$ and Ce$_{3}$Co$_{4}$Sn$_{13}$ is much weaker than their isostructural compounds Sr$_{3}$Ir$_{4}$Sn$_{13}$\cite{PhysRevB.90.035115} and Sr$_{3}$Rh$_{4}$Sn$_{13}$\cite{arXiv:1609.04206}. It is known that the specific heat at low temperatures can be approximately written by $C_{P}/T=\gamma+2.4\pi^{4}nN_{A}K_{B}(/1T_{D}^{3})T^{2}$, where $\gamma, n, N_{A}, K_{B}, T_{D}$ are the Sommerfeld coefficient, free carrier density, Avogadro constant, Boltzmann constant and the Debye temperature, respectively. The first and second terms represent the contributions of itinerant charge carriers and phonons, separately. The obtained Sommerfeld coefficient for La$_{3}$Co$_{4}$Sn$_{13}$ is about 40 $mJK^{2}mol^{-1}$. The value is approximately consistent with 55 $mJK^{-2}mol^{-1}$ obtained from density functional calculations\cite{PhysRevB.79.094424} and reported experimental value of 36 $mJK^{-2}mol^{-1}$ \cite{Israel2005251,LyleThomas20061642,ISI:000340983800029}. The measurements indicate moderate enhancement of carrier effective mass for La$_{3}$Co$_{4}$Sn$_{13}$. We note that the value is also similar to some intermediate valence materials, such as $YbAl_{3}$ \cite{ISI:000086570200314}, $YbFe_{2}Al_{10}$ \cite{arXiv:1612.03421v1}, and Ce$_{3}$Ru$_{4}$Sn$_{13}$ \cite{PhysRevB.94.235151}. However, for Ce$_{3}$Co$_{4}$Sn$_{13}$, we can see a sharp upturn of the $C_{P}/T$ roughly below 3K, which is significantly different from La$_{3}$Co$_{4}$Sn$_{13}$. The behavior at very low temperature can be ascribed to the hybridization effect between conduction electrons and the Ce 4f electrons. The trend of the data are also similar to the reported data in literature \cite{ISI:000340983800029}. In earlier study, the specific heat coefficient \emph{C}(T)/T reaches to a maximum value of about 4 $JK^{-2}mol^{-1}$ at about 0.7 K. This temperature is far below the lowest measurement temperature of present study. The behavior was explained as due to the short-range magnetic correlation of Ce 4f electrons, from which a Kondo temperature of 1.5 K was derived \cite{ISI:000340983800029}.

\begin{figure}[htbp]
  \centering
  \includegraphics[width=7.5cm]{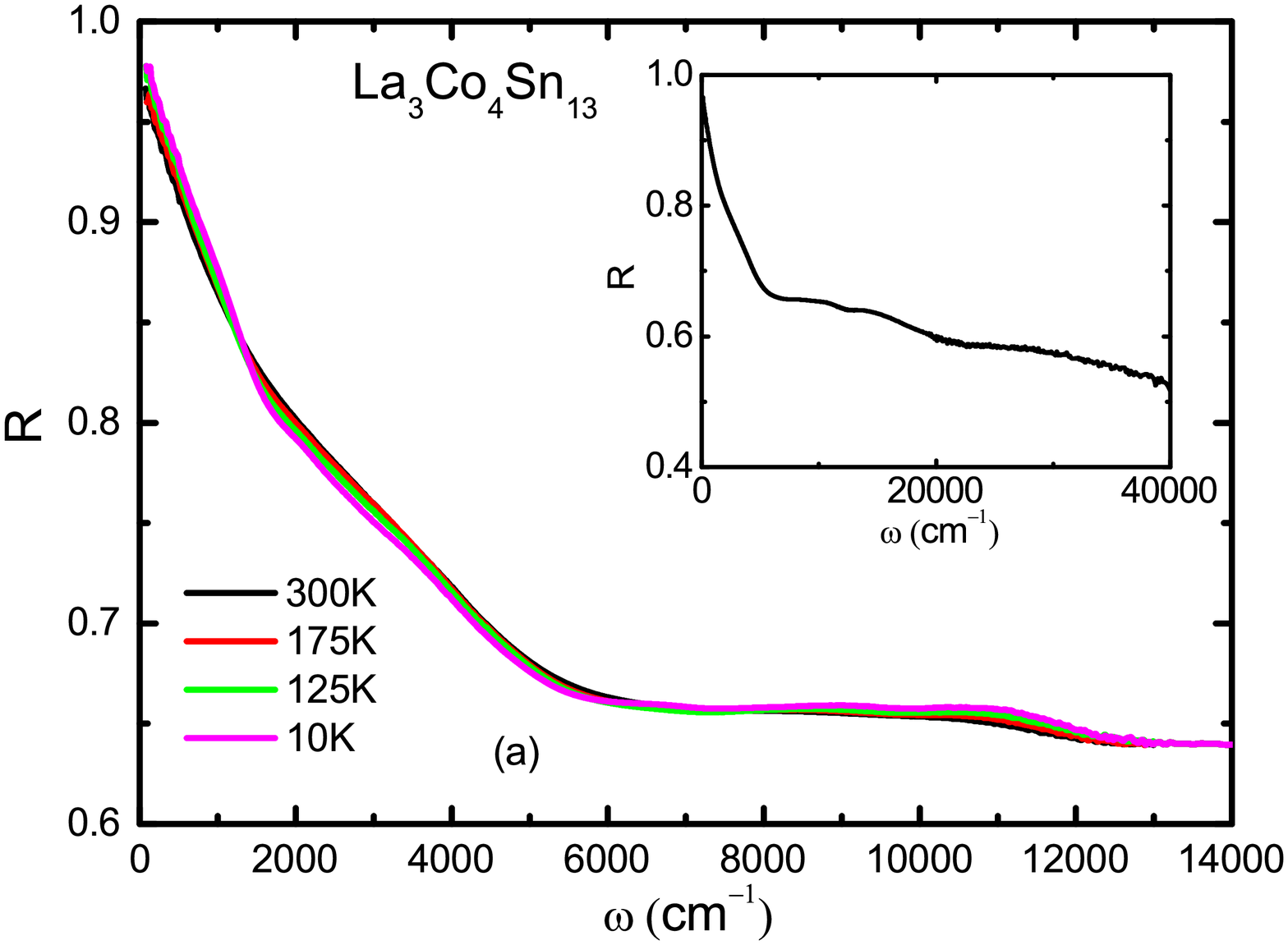}\\
  \includegraphics[width=7.5cm]{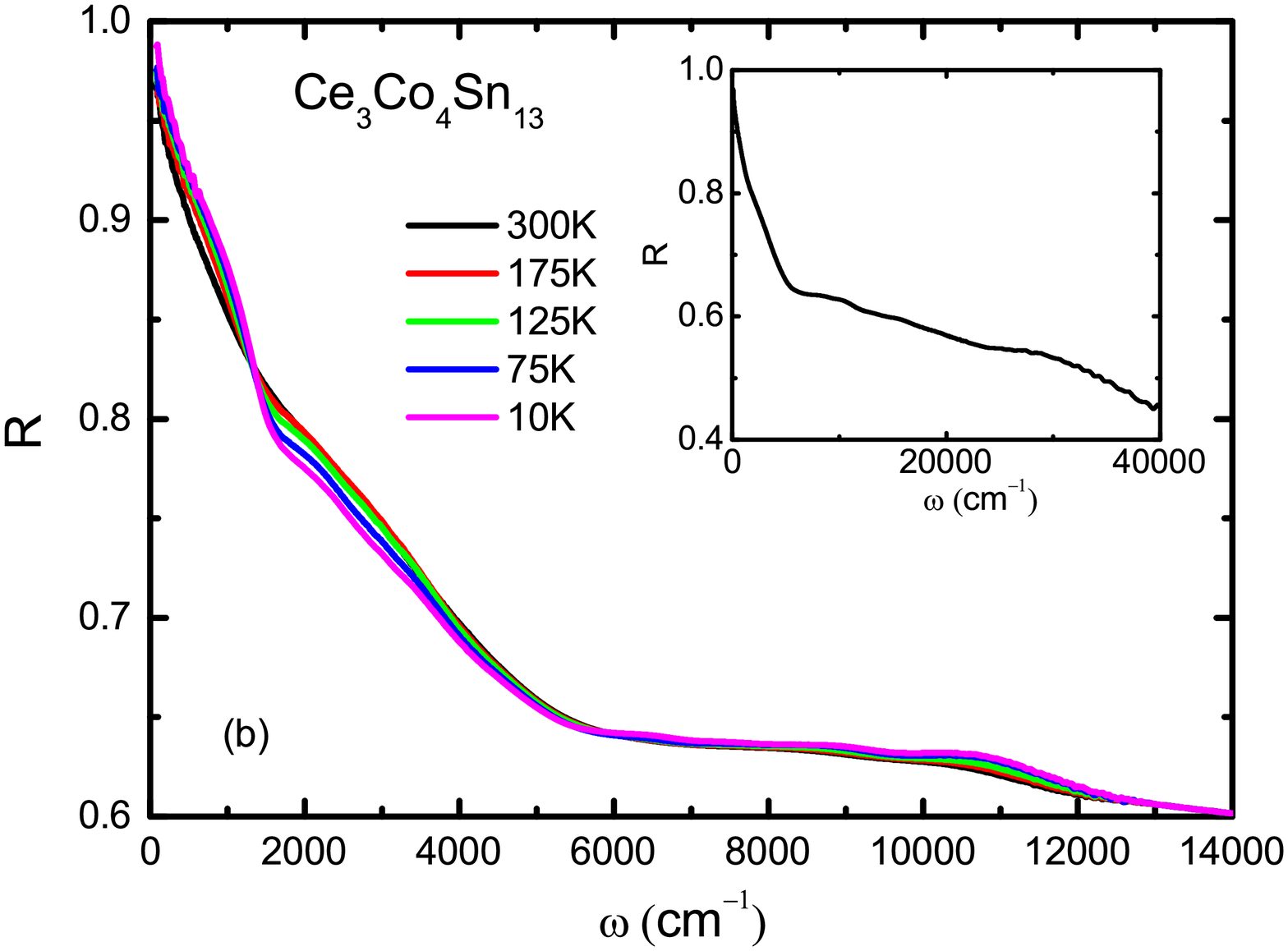}\\
    \caption{The temperature dependent reflectance of La$_{3}$Co$_{4}$Sn$_{13}$ (a) and Ce$_{3}$Co$_{4}$Sn$_{13}$ (b). The inset shows R($\omega$)
data up to 25 000 \cm at 300 K.}\label{Fig:ref}
\end{figure}

Figure \ref{Fig:ref} shows the reflectance spectra R($\omega$) of La$_{3}$Co$_{4}$Sn$_{13}$ and Ce$_{3}$Co$_{4}$Sn$_{13}$ over a broad energy scale at selected temperatures. As can be seen, the values of R($\omega$) approach unity at zero frequency at all temperature, indicating that they both are metallic in nature. By lowering the temperature, we do not see any sharp changes across the structural phase transition temperature $T^{*}$. For Ce$_{3}$Co$_{4}$Sn$_{13}$, with decreasing the temperature, the R($\omega$) between 1000 and 6000 \cm are somewhat suppressed while R($\omega$) between 6000 and 12000 \cm are slightly enhanced. The most prominent phenomenon is that the R($\omega$) develops a dip near 1500 - 2000 \cm, which becomes more and more obvious as the temperature decreasing. The spectral changes may signal the opening of a gap-like feature, as we shall see more clearly in the optical conductivity spectra below. Similar spectral evolutions are also seen in La$_{3}$Co$_{4}$Sn$_{13}$ sample although the suppression features are somewhat weaker.

\begin{figure}[htbp]
  \centering
  \includegraphics[width=7.5cm]{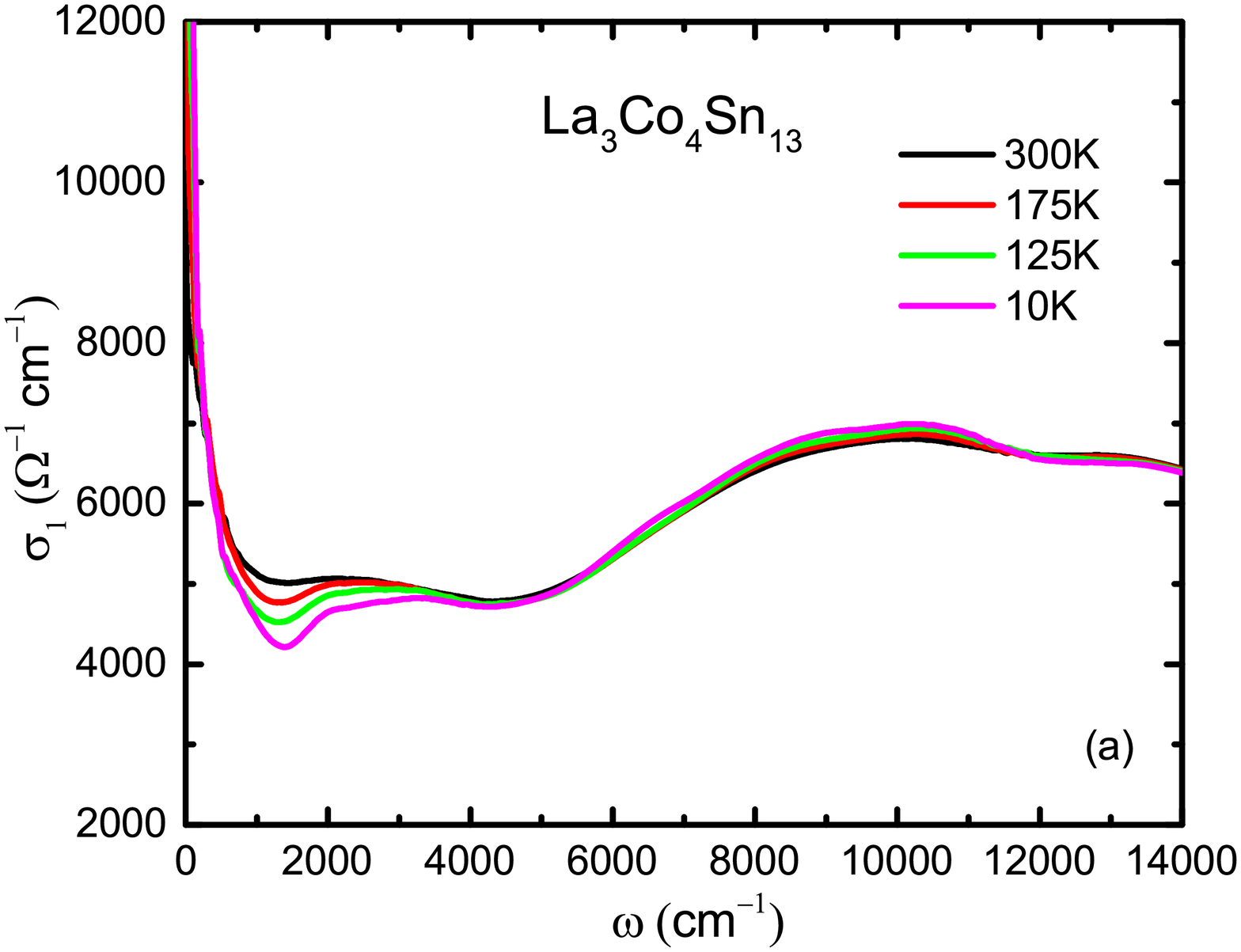}\\
  \includegraphics[width=7.5cm]{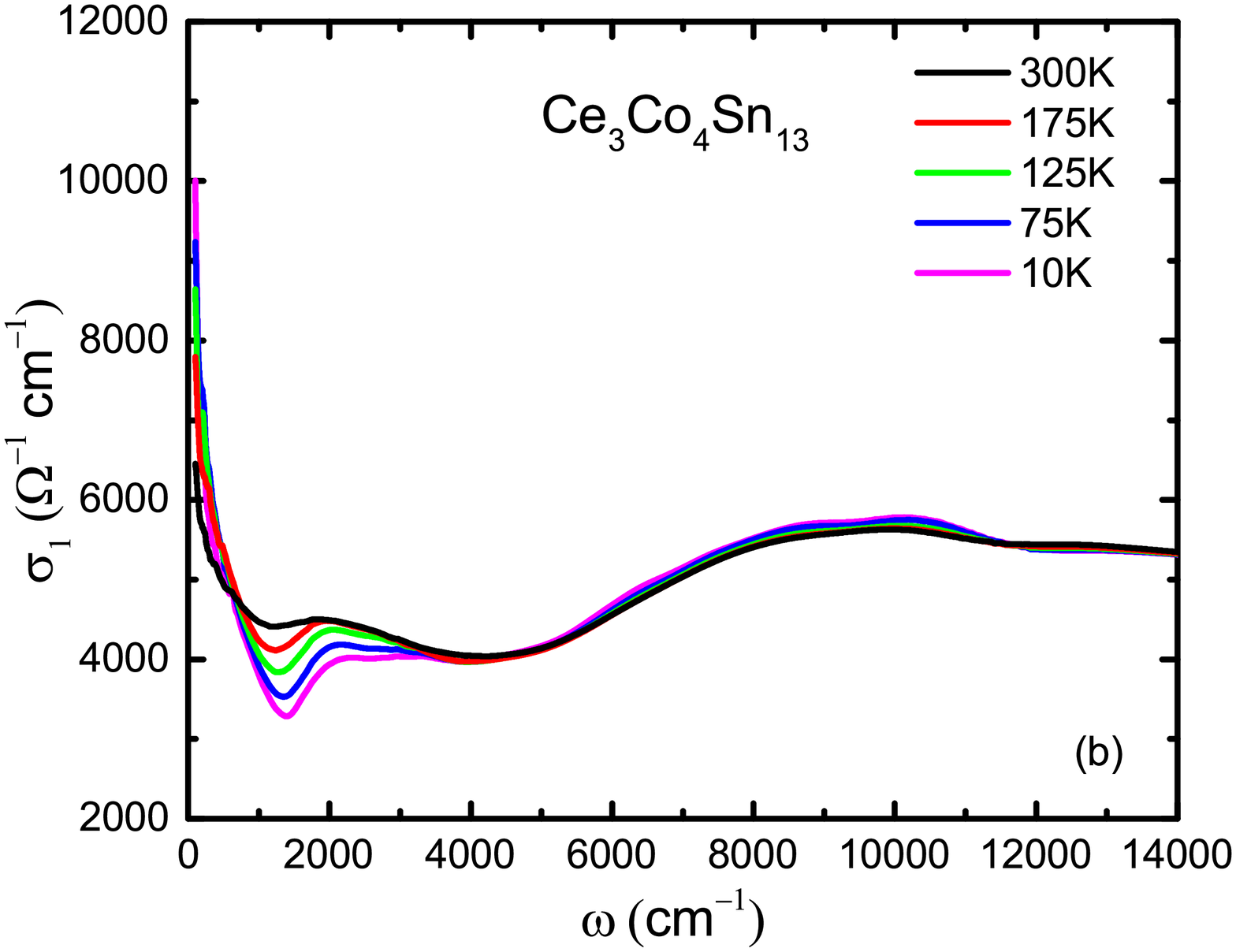}\\
  \caption{The temperature dependent optical conductivity $\sigma_1(\omega)$ of La$_{3}$Co$_{4}$Sn$_{13}$ (a) and Ce$_{3}$Co$_{4}$Sn$_{13}$ (b). The Drude component becomes narrowed at lower temperature.}\label{Fig:conductivity}
\end{figure}

Figure \ref{Fig:conductivity} illustrates the real part of conductivity obtained by the Kramers-Kronig transformation of R($\omega$) at selected temperatures. The evolution of the electronic states is more clearly reflected in the conductivity spectra. The Drude-type conductivity is observed at all temperatures at low frequency for both La$_{3}$Co$_{4}$Sn$_{13}$ and Ce$_{3}$Co$_{4}$Sn$_{13}$, indicating their good metallic response. For Ce$_{3}$Co$_{4}$Sn$_{13}$, upon cooling the optical conductivity between 600 and 4000 \cm is gradually suppressed, and the Drude component becomes narrower. The suppressed spectral weight is transferred mostly to the frequency region between 6000 and 12000 \cm. Similar but relatively weak features are also seen in La$_{3}$Co$_{4}$Sn$_{13}$ sample. The gradual removal of the spectral weight near 2000 \cm in $\sigma_1(\omega)$ represents the progressive formation of a pseudogap feature. The phenomenon is not related to the Ce 4f electrons since the effect is seen in both compounds.

\begin{figure}[htbp]
  \centering
  \includegraphics[width=7.5cm]{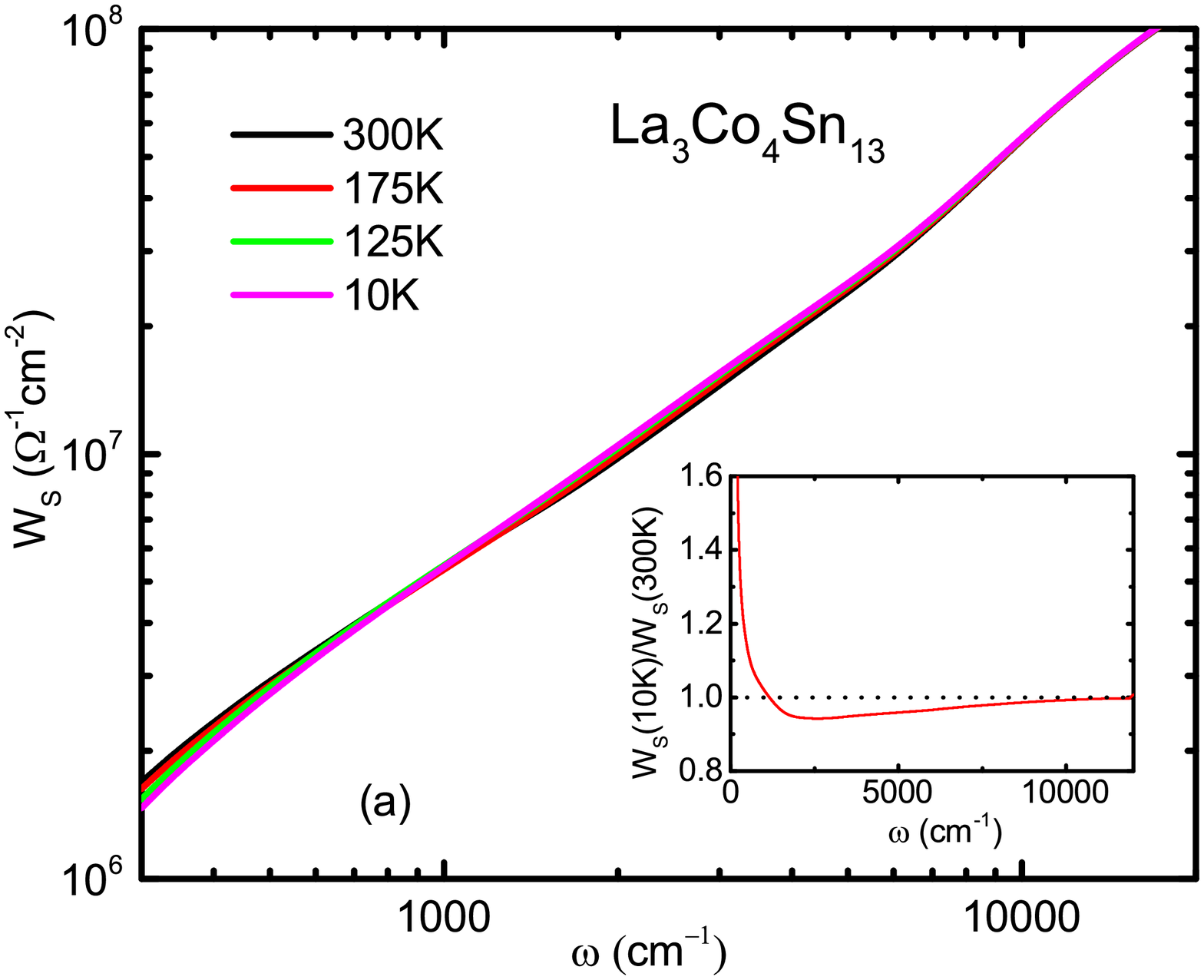}\\
  \includegraphics[width=7.5cm]{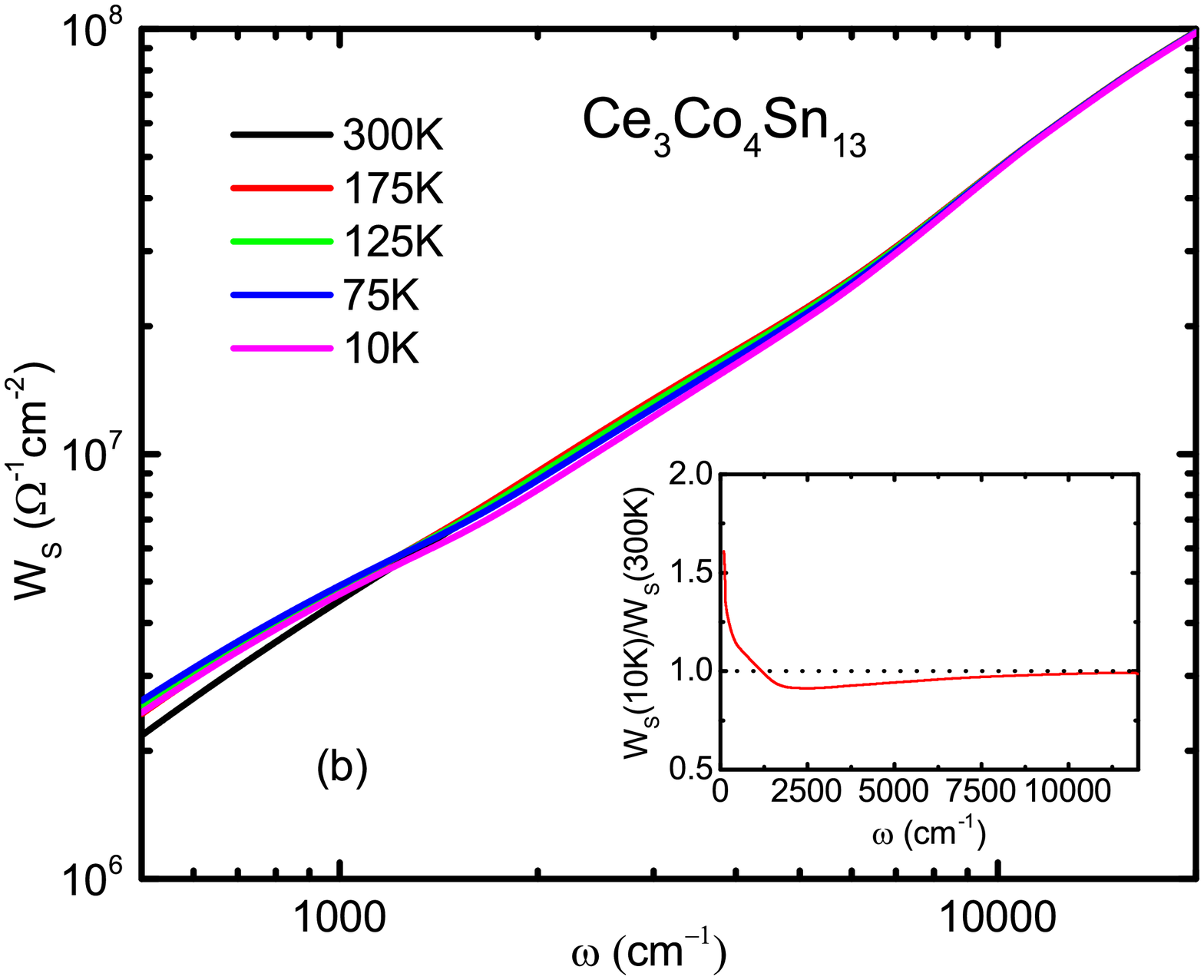}\\
  \caption{Cutoff frequency dependent spectral weight at selected temperatures of La$_{3}$Co$_{4}$Sn$_{13}$ (a) and Ce$_{3}$Co$_{4}$Sn$_{13}$ (b). Inset: the normalized spectral
weight $W_{s}(10 K)/W_{s}(300 K)$ up to 10 000 \cm}\label{Fig:sum-rule}
\end{figure}

To characterize the temperature-induced spectral change, we plot the frequency-dependent integrated spectral weight at different temperatures in Fig. \ref{Fig:sum-rule}. The spectral weight is defined as $W_{S}=\int_{0}^{\omega_{c}}\sigma_{1}(\omega)d\omega$, where $\omega_{c}$ is a cut off frequency. We can see that the spectral weight is gradually recovered at high frequencies. The spectral weight transfer is seen more clearly in the plot of the ratio of the spectral weight at two different temperatures $W_{s}(10 K)/W_{s}(300 K)$, as
shown in the inset of Figure \ref{Fig:sum-rule}. The value of the ratio is higher than unity at very low frequency, which is apparently due to
the higher conductivity values of the narrow Drude component in the low temperature. The ratio becomes less than the unity
at higher energy due to the progressive formation of a gap-like feature with decreasing temperatures. Eventually, the spectral weight is recovered and the ratio approaches unity at higher energies. Such feature is always seen in gap opening systems, such as density wave systems $Ba_{2}Ti_{2}Fe_{2}As_{4}O$ \cite{PhysRevB.90.144508}, $CaFe_{2}(As_{0.935}P_{0.065})_{2}$\cite{ISI:000344013100001}, $ErTe_{3}$\cite{ISI:000296329900003}.

Understanding the origin of the temperature-induced suppression in R($\omega$) and corresponding peak-like feature in $\sigma_1(\omega)$ is of crucial importance for understanding the underlying physics of the two compounds. It is well known that the interband transitions contribute dominantly to the conductivity spectrum at high energies and usually lead peaks in $\sigma_1(\omega)$. However, one would not expect to see a prominent temperature dependence for an interband transition. So this feature near 2000 \cm must have a different origin.

The feature may remind of density wave materials. Because for a density wave order, the opening of an energy gap leads to a spectral weight suppression below $2\Delta$ and a nonsymmetric peak with clear edge-like feature near $2\Delta$ in the optical conductivity\cite{PhysRevLett.76.3838,PhysRevLett.101.257005}. Indeed, the CDW formations below T$^{*}$ were identified based on the specific heat and nuclear magnetic resonance measurements by Lue et al. for both La$_{3}$Co$_{4}$Sn$_{13}$ \cite{PhysRevB.88.115113} and Ce$_{3}$Co$_{4}$Sn$_{13}$ \cite{PhysRevB.85.205120}. But for our optical spectroscopy measurements, the characteristic spectral feature is already present at 300 K. No any other specific structure could be identified across T$^{*}$. Furthermore, for the density wave type energy gap the spectral weight transfer does not extend to very high energy scale. The suppressed spectral weight should be rapidly recovered.
So our finding can not be related to the charge density wave transition. In our earlier studies on isostructural compounds Sr$_{3}$Ir$_{4}$Sn$_{13}$\cite{PhysRevB.90.035115} and Sr$_{3}$Rh$_{4}$Sn$_{13}$\cite{arXiv:1609.04206}, very strong anomalies are present in the specific heat measurements, but we can identify only weak CDW energy gap feature in optical measurement. For the present samples, the phase transition anomaly at $T^{*}$ are much weaker than in the earlier Sr$_{3}$M$_{4}$Sn$_{13}$ (M=Ir, Rh) samples. This may be the reason why we can not identify CDW related features.

The gap-like feature is also commonly observed in heavy fermion materials \cite{PhysRevB.94.035161,0034-4885-79-6-064502,doi:10.7566/JPSJ.85.083702}, originating from the formation of hybridization energy gap in the density of states near the Fermi level. For Ce$_{3}$Co$_{4}$Sn$_{13}$, many experiments had been done to confirm that it is a heavy fermion material \cite{ISI:000340983800029,LyleThomas20061642,ISI:000238426600041,ISI:000254689900072,PhysRevB.86.205113}. However, we noted that essentially the same pseudogap-like feature is also present in La$_{3}$Co$_{4}$Sn$_{13}$, but this compound does not have f electrons. Even though dc resistivity and specific heat measurements indeed reveal some differences between La$_{3}$Co$_{4}$Sn$_{13}$ and Ce$_{3}$Co$_{4}$Sn$_{13}$, for example, the specific heat of Ce$_{3}$Co$_{4}$Sn$_{13}$ begins to increase sharply below about 3 K, an effect caused by the additional f-electron in Ce atoms, the temperature is much lower than the lowest optical measurement temperature. Therefore, in our opinion, the f-electrons have negligible contribution to the pseudogap-like spectral feature.

On the basis of above discussions, we have to attribute the temperature-induced spectral feature to the presence of Co 3d electrons in the compounds. The correlation effect of the 3d electrons leads to the spectral weight transfer or the formation of pseudogap-like feature. The same effect also causes the moderate mass enhancement observed in the specific heat measurement. In fact, Kondo latttice formation or heavy fermion behavior have been observed in the d-electron systems, for example in $LiV_{2}O_{4}$ \cite{doi:10.7566/JPSJ.85.091009,Das2009Absence,Anisimov1999Electronic,J2007Correlation,Irizawa2011Direct,Fujimoto2002Hubbard,Kadono2012Quasi,Tomiyasu2014Spin}. It was proposed that the mechanism for the heavy quasi-particle formation in $LiV_{2}O_{4}$ is similar to the f-electron systems, i.e., hybridization between the ¡®¡®localized¡¯¡¯ $A_{1g}$ band and the ¡®¡®itinerant¡¯¡¯ $E_{g}$ band arising from the combination effect of crystal field splitting and d-d Coulomb interaction \cite{Anisimov1999Electronic}. This picture may also apply to the present study. In R$_{3}$Co$_{4}$Sn$_{13}$ (R=La, Ce), the Co atoms form the Co(Sn2)$_6$ trigonal prisms which are corner sharing with a titled three-dimensional (3D) arrangement. As we mentioned in the introduction, under the Co(Sn2)$_6$ trigonal prism crystal field, the 3d orbits of Co are split into $d_{z^{2}}$, degenerate $d_{xy}(d_{x^{2}+y^{2}})$, and $d_{xz}(d_{yz})$ orbits from low to high energy. The two antibonding bands mixed by Co $d_{xz}(d_{yz})$ and Sn2 5p cross the $E_{F}$ to form a relatively broad conduction band and the complicated Fermi surface \cite{PhysRevB.79.094424}. The $d_{z^{2}}$ and degenerate $d_{xy}(d_{x^{2}+y^{2}})$ bands are localized. The d-d Coulomb interaction was indicated in earlier x-ray photoemission spectroscopy study for La$_{3}$Co$_{4}$Sn$_{13}$ and Ce$_{3}$Co$_{4}$Sn$_{13}$ \cite{PhysRevB.91.035101}. Therefore, we speculate the hybridization may exist between itinerant Co $d_{xz}(d_{yz})$ or Sn2 5p bands and other localized 3d bands, which is similar as the case of $LiV_{2}O_{4}$, although a detailed picture remains to be established for the formation of such gap-like lineshape. The spectral weight transfer could be related to the energy scale of the renormalized d-d Coulomb interactions.

\section{conclusion}

In summary, we have successfully grown single-crystal samples of La$_{3}$Co$_{4}$Sn$_{13}$ and Ce$_{3}$Co$_{4}$Sn$_{13}$ and conducted careful characterizations by electrical resistivity and specific heat measurements. They both exhibit weak phase transitions approximately at 150K. La$_{3}$Co$_{4}$Sn$_{13}$ shows a superconducting transition near 3 K and specific heat measurement indicates a moderate mass enhancement. However, Ce$_{3}$Co$_{4}$Sn$_{13}$ shows very strong enhancement of specific heat below about 4 K arising from the Ce 4f electrons. We performed optical spectroscopy measurements on both samples. We do not find density wave type energy gap formation across the phase transition near 150 K. Instead, we observe clear temperature-dependent spectral weight suppression below about 4000 \cm in conductivity spectra $\sigma_1(\omega)$ for both compounds, indicating the progressive formation of gap-like features with decreasing temperature. The suppressed spectral weight transfers mostly to the higher energy region. The observation reflects the presence of correlation effect in the compounds. We elaborate that the correlation effect should come from the transition metal Co 3d electrons.

\begin{center}
\small{\textbf{ACKNOWLEDGMENTS}}
\end{center}

This work was supported by the National Science Foundation of China (No. 11327806, GZ1123) and the National Key Research and Development Program of China (No.2016YFA0300902).

\bibliographystyle{apsrev4-1}
  \bibliography{La3Co4Sn13}

\end{document}